 \documentclass[final,authoryear,3p,times,twocolumn]{elsarticle}
\usepackage{amssymb}
\journal{}

\begin{document}

\begin{frontmatter}

%% Title, authors and addresses

%% use the tnoteref command within \title for footnotes;
%% use the tnotetext command for the associated footnote;
%% use the fnref command within \author or \address for footnotes;
%% use the fntext command for the associated footnote;
%% use the corref command within \author for corresponding author footnotes;
%% use the cortext command for the associated footnote;
%% use the ead command for the email address,
%% and the form \ead[url] for the home page:
%%
%% \title{Title\tnoteref{label1}}
%% \tnotetext[label1]{}
%% \author{Name\corref{cor1}\fnref{label2}}
%% \ead{email address}
%% \ead[url]{home page}
%% \fntext[label2]{}
%% \cortext[cor1]{}
%% \address{Address\fnref{label3}}
%% \fntext[label3]{}

\title{A New Galactic Wolf-Rayet Star in Centaurus}

%% use optional labels to link authors explicitly to addresses:
%% \author[label1,label2]{<author name>}
%% \address[label1]{<address>}
%% \address[label2]{<address>}

\author{A. Roman-Lopes}

\address{Physics Department - Astronomy Group - Universidad de La Serena, Cisternas 1200,
    La Serena, Chile}

\begin{abstract}

In this work I communicate the detection of a new Galactic Wolf-Rayet star (WR60a) in Centaurus.
The H- and K-band spectra of WR60a, show strong carbon near-infrared emission lines, characteristic of Wolf-Rayet stars of the WC5-7 sub-type. 
Adopting mean absolute magnitude M$_K$ and mean intrinsic ($J-K_S$) and ($H-K_S$) colours, it was found that WR60a suffer a mean visual extinction 
of 3.8$\pm$1.3 magnitudes, being located at a probable heliocentric distance of 5.2$\pm$0.8 Kpc, which for the related Galactic longitude (l=312) 
puts this star probably in the Carina-Sagittarius arm at about 5.9 kpc from the Galactic center. 
I searched for clusters in the vicinity of WR60a, and in principle found no previously known clusters in a search radius region of several tens arc-minutes.
The detection of a well isolated WR star induced us to seek for some still unknown cluster, somewhere in the vicinity of WR60a.
From inspection of 5.8$\mu$m and 8.0$\mu$m Spitzer/IRAC GLIMPSE images of the region around the new WR star, it was found strong mid-infrared extended
emission at about 13.5 arcmin south-west of WR60a.
The study of the the H-K$_S$ colour distribution of point sources associated with the extended emission, reveals the presence of a 
new Galactic cluster candidate probably formed by at least 85 stars.

\end{abstract}

\begin{keyword}
%% keywords here, in the form: keyword \sep keyword
stars: Wolf-Rayet; Galaxy: stellar content
%% MSC codes here, in the form: \MSC code \sep code
%% or \MSC[2008] code \sep code (2000 is the default)

\end{keyword}

\end{frontmatter}

% \linenumbers

%% main text

\section{Introduction}

Despite the important role that WR stars probably play in shaping both, Galactic structure and chemical evolution, the total number of 
WR stars in the Milk Way is still an open issue. \citet{b6,b7} estimate a number between 1000-2500 WR stars
to be located within the Galaxy, which 
compared to the about 300 known Galactic WR stars \citep {b13}) indicates that up to date, there still a large number of such stars to be
discovered in our Galaxy.  
One reason for this comes from the fact that dust obscuration makes hard to observe WR stars in the optical window
through the entire Galactic plane. 
On the other hand, infrared observations can provide the means for finding a significant fraction of the remaining population of obscured 
Galactic WR stars. 
Indeed, in the Near-Infrared (NIR) window, $H$ and $K$ band spectroscopy may be used to properly identify and classify the new Galactic WR stars, 
accordingly to the known WC, WN and WO subtypes \citep{b1,b2,b3}.

In this work I report the discovery of a new Wolf-Rayet star in Centaurus through the detection of its NIR carbon and He emission lines, using 
ESO-NTT-SOFI archival spectroscopic data.
In section 2 I describe the observational data and the data reduction procedures, in section
3 I present the results, and in section 4 it is presented a summary of the work.

\section{Observational data and reduction procedures}

\subsection{Photometric parameters of the newly-identified WR star}

Coordinates and photometric parameters for the new WR star are shown in
Table 1. The near-IR photometric data were taken from the Two-Micron All Sky Survey
2MASS, \citep{b14} using the
NASA/IPAC\footnote{http://irsa.ipac.caltech.edu/applications/BabyGator/}
Infrared Science Archive, while the V-band photometry was obtained
from the Naval Observatory Merged Astronomical Dataset
(NOMAD\footnote{http://www.nofs.navy.mil/data/fchpix/vo\_nofs.html}). 
The star was named WR60a, following the common practice of to
give numbers to Galactic WR stars in RA order, with further additions  
between integers from \citet{b7,b13}.
A 2MASS $K_S$ band finding chart for this object is presented in Figure 1.

%Table 1

\begin{table*}
%\begin{minipage}[t]{\columnwidth}
\caption{Some photometric parameters of the WR60a star.
The near-IR photometric data were taken from the Two-Micron All Sky Survey
(2MASS) using the NASA/IPAC
Infrared Science Archive, while the V-band photometry was obtained
from the Naval Observatory Merged Astronomical Dataset (NOMAD).
For the derived spectral type, visual extinction and heliocentric distance, see the text.}
%\vspace{10}
\label{catalog}
\centering
\renewcommand{\footnoterule}{}  % to avoid a line before footnotes
\begin{tabular}{cccccccccc}
\hline \hline
ID & RA     &  Dec  & $V$  & $J$ & $H$ & $Ks$ & Sp type & A$_V$ (mag) & d (Kpc) \\
 &(J2000) & (J2000) &    & \\
\hline
  WR60a  &14:06:03.61  & -60:27:29.5  &   15.35  &   10.84  & 10.12 & 9.39 & WC5-7 & 3.8$\pm$1.3 & 5.2$\pm$0.8 \\
\hline
\end{tabular}
%\end{minipage}
\end{table*} 

%% In a manner similar to \objectname authors can provide links to dataset
%% hosted at participating data centers via the \dataset{} command.  The
%% second curly bracket argument is printed in the text while the first
%% parentheses argument serves as the valid data set identifier.  Large
%% lists of data set are best provided in a table (see Table 3 for an example).
%% Valid data set identifiers should be obtained from the data center that
%% is currently hosting the data.
%%
%% Note that AASTeX interprets everything between the curly braces in the 
%% macro as regular text, so any special characters, e.g. "#" or "_," must be 
%% preceded by a backslash. Otherwise, you will get a LaTeX error when you 
%% compile your manuscript.  Special characters do not 
%% need to be escaped in the optional, square-bracket argument.

\subsection{Near-Infrared spectroscopic observations}

%Figure1
   \begin{figure}
    %\vspace{302pt}
   \centering
   \includegraphics[bb=14 14 337 298,width=7 cm,clip]{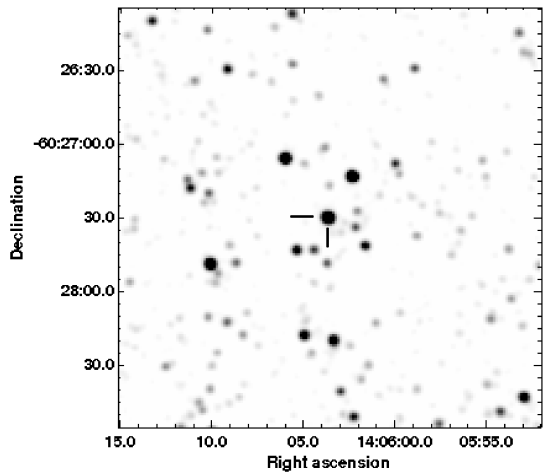}
      \caption{The 4 arcmin $\times$ 4 arcmin 2MASS $K_S$ band finding chart for the newly-discovered Wolf-Rayet star WR60a. 
North is to the top, East to the left.}
         \label{FigVibStab}
   \end{figure} 

In this work I used ESO\footnote{http://archive.eso.org/eso/eso\_archive\_main.html} archival data obtained 
with the SofI instrument \citep{b8}, coupled to the 3.6m NTT telescope.
The spectra were taken as part of the ESO program 075.D-0210(A) (PI A.P. Marston and collaborators), with the targets being selected
accordingly the technique presented by \citet{b4}. A summary of the spectroscopic dataset is presented in Table 2.

\subsection{Data reduction}

The raw spectra were reduced following the NIR reduction procedures presented by \citet{b9}, shortly describe here.
The two-dimensional frames were sky-subtracted for each pair of
images taken at the two nod positions A and B, followed by division
of the resultant image by a master flat. 
The multiple exposures were combined, followed by one-dimensional extraction of the spectra.
Thereafter, wavelength calibration was applied using the sky lines;
the typical error (1-$\sigma$) for this calibration was $\sim$21~\AA.
Telluric atmospheric correction done using $H$ and $K$ band spectra of A and B type stars, completed the
reduction process.

%Figure2
   \begin{figure*}
    %\vspace{302pt}
   \centering
   \includegraphics[bb=14 14 415 255,width=10 cm,clip]{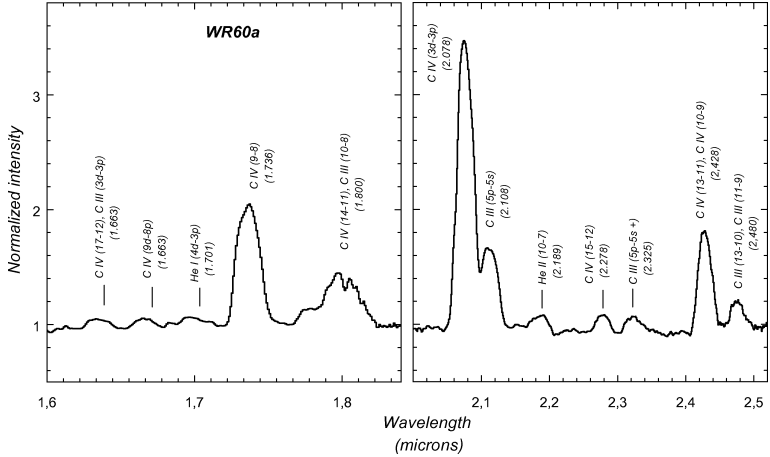}
      \caption{The continuum normalized $H$ and $K$ band spectra of the new Wolf-Rayet star WR60a.}
         \label{FigVibStab}
   \end{figure*} 

\section{Results}

\subsection{The H- and K-band spectra of WR60a}

The H- and K-band spectra of WR60a are shown in Figure 2, with the main 
helium and carbon emission lines identified by labels.
From an inspection of the this figure we can see that the carbon emission lines are very strong,
with the most prominent being the C{\sc iv} line at 2.076{$\mu$m}.
On the other hand, from a comparison of the WR60a K-band spectrum with those from the spectral atlas of WRs stars of \citet{b10}, we might 
conclude that this new WR star is probably of the WC5-7 subtype. Indeed, from the ratio (3.7) of the C{\sc iv} 2.076$\mu$m and 
C{\sc iii} 2.110$\mu$m equivalent widths (EWs), one can get the same conclusion \citep{b12}.  

\subsubsection{A probable non-dusty WC star}

Late-type WC stars sometimes are surrounded by circumstellar dust (which is believed to be produced by interaction with an
OB binary companion), whose thermal emission may contribute to the underlying NIR continuum. As can be noticed in Figure 2, 
the K-band carbon lines of WR60a do not seem to suffer such effect in its underlying continuum.
In one hand, the observed NIR colours of WR60a are compatible with that for a WC5-7 WR star ((J-K$_S$)$_0$=0.62, (H-K$_S$)$_0$=0.58 - \citet{b12}), 
reddened by 3.8$\pm$1.3 mag in the V band \citep{b15}.
On the other hand, the absence of reduced line strengths in principle 
can not be used to discard the presence of a line-of-sight/binary early-type companion since relatively short period late-type WC binaries
like $\gamma$ Vel (WC8+O), are known not to form dust.

%Table 2

\begin{table}
%\begin{minipage}[t]{\columnwidth}
\caption{Summary of the NTT/SofI dataset used in this work.}
\label{catalog}
\centering
\renewcommand{\footnoterule}{}  % to avoid a line before footnotes
\begin{tabular}{cc}
%\hline \hline
%ID & SpType \\
\hline
   Date  & 2005-06-23\\
   Telescope  & NTT\\
   Instrument & SofI\\
   Grism  & GR\\
   Slit (arcsec) & 0.6 x 290\\
   Resolution  & 1000\\
   Coverage ($\mu$m) & 1.53-2.52\\
   Seeing (arcsec)  & 1-2\\
\hline
\end{tabular}
%\end{minipage}
\end{table} 

\subsubsection{Heliocentric distance to WR60a}

Despite quantitative analysis of the newly-discovered WR star is beyond the scope of this work,
from the derived spectral type and associated NIR photometry, it is possible to obtain an estimate for the heliocentric distance to 
WR60a, considering the extinction law for NIR bands derived by \citet{b15} and following the procedure described by \citet{b3}.
I used the mean values for absolute K-band magnitude (M$_K$=-4.59), and J-K$_S$ and H-K$_S$ colours (0.62 and 0.58, respectively) taken from \citet{b12}.
It was found a mean heliocentric distance of 5.2$\pm$0.8 kpc, which puts this WR star at about 5.9 
kpc from the Galactic center, probably in the Carina-Sagittarius arm.

\subsection{Cluster candidates and other WR stars in the vicinity of WR60a}

I searched for cataloged clusters in the vicinity of WR60a, and found no known clusters 
around it in a region of tens of arc-minutes. 
In fact, the first known cluster is DBS2003 \#135 \citep{b11} with coordinates 14:08:42 -61:10:36, and located at about 47 arcmin 
from the new WR star.
Also from a search in the clusters candidate list of \citet{b16}, the closest one is their number \#43 ($\alpha$=14:00:28, 
$\delta$=-60:59:15 - J2000) that is located at about 52 arcmin from WR60a.

The presence of a well isolated WR star naturally generates the question about the WR60a parental cluster candidate. 
In fact, the presence of such kind of star strongly suggest the existence of some nearby unknown stellar cluster. 
In order to address this question, I performed a visual search through 2MASS and Spitzer/IRAC images of the region towards the new WR star.
I found no indication for the presence of clusters from a visual inspection of the 2MASS images. However, from the 5.8$\mu$m and 8.0$\mu$m 
Spitzer/IRAC images it can be noticed the presence of strong extended emission, south-west of WR60a. In Figure 3 I present a false-color image
made using the 4.5$\mu$m, 5.8$\mu$m and 8.0$\mu$m Spitzer/IRAC images of this Galactic sector, taken by the GLIMPSE survey. There, it is clearly
seem the presence of several filamentary structures distributed perpendicularly to the Galactic plane, and extending at least to b$\sim$1.1.
On the other hand, an inspection of the filamentary region indicates the existence of an apparent concentration of near- to mid-infrared point
sources around the coordinates (J2000) 14:05:05.42, -60:31:22.9, and 14:05:35.07,
-60:40:30.5 (hereafter named cluster candidate \#1 and cluster candidate \#2 - see Figure 3), both located at angular distances of 
8 arcmin and 13.5 arcmin, respectively from WR60a. 

Finally, I also searched for other known Wolf-Rayet stars in the vicinity of WR60a, however only two were found at
angular distances greater than a degree! HD121194 (WC7) with coordinates 13:55:48.1, -61:09:50 (J2000), and SSTGLMCG313.8558+00.6489 (WN6) with
coordinates 14:21:23.15, -60:18:04.1 (J2000).

%Figure3
   \begin{figure}
    %\vspace{302pt}
   \centering
   \includegraphics[bb=14 14 535 598,width=8 cm,clip]{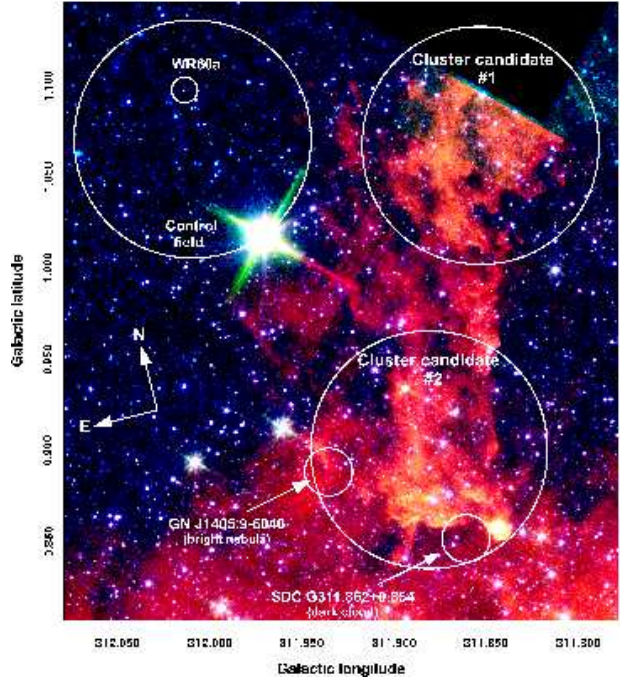}
      \caption{The Spitzer/IRAC 4.5$\mu$m (blue), 5.8$\mu$m (green) and 8.0$\mu$m (red) false-color image of the region around WR60a.}
         \label{FigVibStab}
   \end{figure} 

\subsubsection{Known astrophysical sources possible associated to the cluster candidates \#1 and \#2}

Using the SIMBAD Astronomical Database I took a look in the literature for astrophysical sources around the central 
coordinates of both cluster candidates, and found no sources related to the cluster candidate \#1.
On the other hand, for the cluster candidate \#2, there are two objects found at about 3 arcmin from its assigned coordinates. 
The first measuring about 1.7 arcmin $\times$ 1.1 arcmin is the bright nebula named GN J1405.9-6040 (14:05:58.53, -60:40:14.4 (J2000)),
cataloged by \citet{b17} from inspection of DSS-2 blue and red images. The second is the dark nebula SDC G311.862+0.854 (14:05:28.32, -60:43:37 (J2000)), 
which was cataloged by \citet{b18} in a work aimed to identify Galactic infrared dark clouds (IRDCs), using Spitzer GLIMPSE and MIPSGAL data.
The relative positions of this objects are indicated in Figure 3. It is interesting to notice that both nebula are
placed at the border of the region assigned to the cluster candidate \#2, which may be indication 
of ongoing star formation within the host molecular clouds.
However, only the detection of extended mid-infrared emission is not enough to properly support the assumption of the presence of a cluster of stars
there. In the next section, it is presented some results that favor this assumption for the cluster candidate \#2.

\subsubsection{NIR Colour study of the stellar population associated to the cluster candidates \#1 and \#2}

%Figure4
   \begin{figure}
    \vspace{-80pt}
   \centering
   \includegraphics[bb=14 14 235 523,width=5 cm,clip]{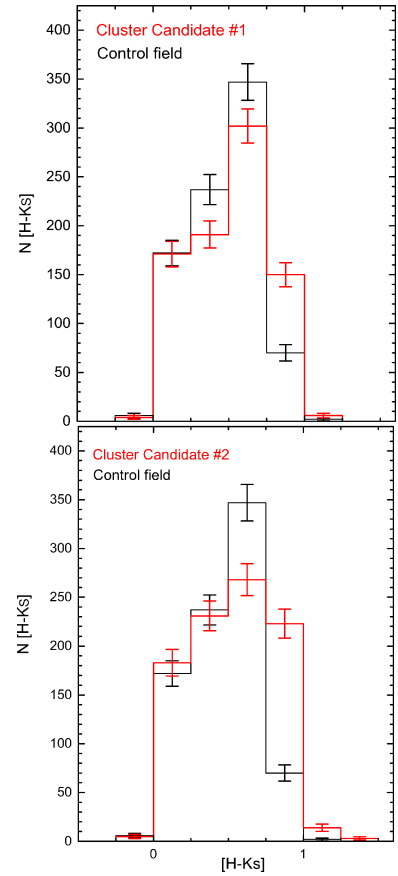}
      \caption{The H-K$_S$ colours histograms for the 2MASS sources detected in the direction of the two cluster candidates, together with that
obtained from the control field (see Figure 3).}
         \label{FigVibStab}
   \end{figure} 

The detailed study of the stellar population associated to the two cluster candidates identified in the previous section, is out of the 
scope of this work. However, even a simple analyses of the NIR colour distributions can provide additional clues to support (or not) the 
idea of existence of clusters of stars associated with the two selected regions in Figure 3. In this sense, it was analyzed the H-K$_S$ colour 
distributions of point sources from three delimited areas, labeled \textit{cluster candidate \#1},
\textit{cluster candidate \#2} and \textit{control field}, the last centered at coordinates (J2000) 14:06:06.57, -60:28:46.2 and assumed 
to be mainly compound by Galactic field stars (with the obvious exception of WR60a). 
The study was done using the 2MASS H- and K$_S$-band photometry of sources present in the three selected regions, considering circular 
fields of 4 arcmin radius (as illustrated in Figure 3 by the white large circles). 
The results are shown in Figure 4.
From these diagrams we can see that the stellar population in the direction of the cluster candidate \#1 is mainly formed 
by Galactic field stars. In fact, its N(H-K$_S$) $\times$ H-K$_S$ diagram presents an excess of field sources in the colour 
range 0.25$\leq $H-K$_S$$\leq$0.75 (91$\pm$32) that is fully compensated by the excess of cluster \#1 sources in the range 
0.75$\leq $H-K$_S$$\leq$1.0 (81$\pm$14).
On the other hand, the N(H-K$_S$) $\times$ H-K$_S$ diagram for sources in the cluster candidate \#2 region, clearly indicates an excess of objects
with H-K$_S$ colours greater than 0.75 mag. Indeed, the number of extra sources found in the control region in the colour range 0.5$\leq $H-K$_S$$\leq$0.75 
(80$\pm$25) is much less than that computed for the colour range 0.75$\leq $H-K$_S$$\leq$1.25 (165$\pm$18). From this results, I can crude estimate as
85$\pm$31 the number of probable members of the cluster candidate \#2. Considering that the photometric completeness of the 2MASS survey
in the H and K$_S$ bands restricts the analyses mainly to the more luminous sources, the true number of cluster members probably is greater than the
inferred here.
Of course, new further spectrophotometric studies are necessary to properly constrain the cluster population in this part of the Galaxy.   

\section{Summary}

In this work the detection of a new Galactic Wolf-Rayet star in the direction of Centaurus is reported. The H- and K-band spectra
of WR60a indicate that the star probably belongs to the WC5-7 subtype.
From the derived spectral type and associated NIR photometry, it was computed a probable heliocentric distance of 5.2$\pm$0.8 Kpc, which 
for the related Galactic longitude (l=312) puts this WR star at about 5.9 kpc from the Galactic center, probably in the Carina-Sagittarius arm.

I searched for clusters in the vicinity of WR60a, and in principle found no known cataloged clusters in a radius search region of several tens of arc minutes. 
On the other hand, I also searched for other known Wolf-Rayet stars in the vicinity of WR60a, with no detections inside an one degree search radius!

The detection of a well isolated WR star naturally leaved us to seek for some still unknown cluster, somewhere in the vicinity of WR60a.
From visual inspection of 5.8$\mu$m and 8.0$\mu$m Spitzer/IRAC GLIMPSE images of the region around the new WR star, it was found strong mid-infrared extended
emission produced by filamentary structures oriented perpendicularly to the Galactic plane. The study of the the H-K$_S$ colour distributions 
of point sources associated with the extended emission, indicates the presence of a new Galactic cluster candidate (central coordinates (J2000) 
14:05:35.07, -60:40:30.5) placed at about 13.5 arcmin south-west of WR60a, for which I estimate a lower-limit of 85$\pm$31
cluster members.

\section{acknowledgments}

      %The author thanks the suggestions and comments made by the anonymous referee. 
      This work was partially supported by the ALMA-CONICYT Fund, under the project number 31060004,
      "A New Astronomer for the Astrophysics Group, Universidad de La Serena", and by the Physics department of the 
      Universidad de La Serena.
      This publication makes use of data products from the Two Micron All Sky Survey, which is a joint project of the University of 
      Massachusetts and the Infrared Processing and Analysis Center/California Institute of Technology, funded by the
      National Aeronautics and Space Administration and the National Science Foundation.
      This research has made use of the SIMBAD database, operated at CDS, Strasbourg, France.
      Based on observations made with ESO Telescopes at the La Silla Observatory under programme ID 075.D-0210(A).

\label{}

\bibliographystyle{elsarticle-harv}
\bibliography{<your-bib-database>}

%% Authors are advised to submit their bibtex database files. They are
%% requested to list a bibtex style file in the manuscript if they do
%% not want to use elsarticle-harv.bst.

%% References without bibTeX database:

% \begin{thebibliography}{00}

%% \bibitem must have one of the following forms:
%%   \bibitem[Jones et al.(1990)]{key}...
%%   \bibitem[Jones et al.(1990)Jones, Baker, and Williams]{key}...
%%   \bibitem[Jones et al., 1990]{key}...
%%   \bibitem[\protect\citeauthoryear{Jones, Baker, and Williams}{Jones
%%       et al.}{1990}]{key}...
%%   \bibitem[\protect\citeauthoryear{Jones et al.}{1990}]{key}...
%%   \bibitem[\protect\astroncite{Jones et al.}{1990}]{key}...
%%   \bibitem[\protect\citename{Jones et al., }1990]{key}...
%%   \harvarditem[Jones et al.]{Jones, Baker, and Williams}{1990}{key}...
%%

% \bibitem[ ()]{}

% \end{thebibliography}

\end{document}